\documentclass[conference]{IEEEtran}
\usepackage[hyphens,spaces,obeyspaces]{url}
\usepackage{acronym}
\usepackage{multirow}
\usepackage{adjustbox}
\usepackage{graphicx}
\usepackage{amsmath}
\usepackage{subcaption}
\usepackage{filecontents}
\usepackage{colortbl}
\usepackage{textcomp} % for arrow in text
\usepackage{enumitem}
\usepackage[ruled]{algorithm2e}
\usepackage{threeparttable}
\usepackage{tablefootnote}
\usepackage{lipsum}
\usepackage{todonotes}
\usepackage[mathlines,switch]{lineno}
\usepackage{pifont}% http://ctan.org/pkg/pifont
\usepackage[english]{babel}
\usepackage{pdfpages}
\usepackage[style=ieee, citestyle=numeric-comp, backend=biber]{biblatex}
\usepackage{setspace}
\begin{document}

\title{Datasheet for Subjective and Objective Quality Assessment Datasets}
\author{
    \IEEEauthorblockN{ Nabajeet Barman\IEEEauthorrefmark{1}\IEEEauthorrefmark{3}, Yuriy Reznik\IEEEauthorrefmark{2}, and Maria Martini\IEEEauthorrefmark{3}}
    \IEEEauthorblockA{\IEEEauthorrefmark{1}Brightcove UK Ltd, London, United Kingdom, nbarman@brightcove.com}
    \IEEEauthorblockA{\IEEEauthorrefmark{2}Brightcove Inc, Seattle, USA, yreznik@brightcove.com}
    \IEEEauthorblockA{\IEEEauthorrefmark{3}Kingston University, London, United Kingdom, m.martini@kingston.ac.uk}
    }
\maketitle

\begin{abstract}
Over the years, many subjective and objective quality assessment datasets have been created and made available to the research community. However, there is no standard process for documenting the various aspects of the dataset, such as details about the source sequences, number of test subjects, test methodology, encoding settings, etc. Such information is often of great importance to the users of the dataset as it can help them get a quick understanding of the motivation and scope of the dataset. Without such a template, it is left to each reader to collate the information from the relevant publication or website, which is a tedious and time-consuming process. In some cases, the absence of a template to guide the documentation process can result in an unintentional omission of some important information. 

This paper addresses this simple but significant gap by proposing a datasheet template for documenting various aspects of subjective and objective quality assessment datasets for multimedia data. The contributions presented in this work aim to simplify the documentation process for existing and new datasets and improve their reproducibility. The proposed datasheet template is available on GitHub\footnote{https://github.com/NabajeetBarman/datasheet-for-qoe-datasets}, along with a few sample datasheets of a few open-source audiovisual subjective and objective datasets.
\end{abstract}


\IEEEpeerreviewmaketitle

\section{Introduction}

Over the past two decades, video streaming has become ubiquitous, with it currently comprising approximately 82\% of total internet traffic \cite{Cisco2018}. This has largely been possible due to the advancements in various aspects of multimedia streaming from improved codecs \cite{h264,h265,h266} to better CDNs, improved transport, and delivery mechanisms \cite{HLS,DASH} to more powerful and high-quality end-user devices such as smartphones, smart TVs, and laptops. However, for the continued growth of such video streaming services delivering multimedia content over the internet, it is important to ensure that the end user is satisfied with the service's quality of experience (QoE).

QoE is defined in ITU-T Rec P.10/G.100~\cite{p10_g100} as ``\textit{The degree of delight or annoyance of the user of an application or service}". Over the years, 
% MM - I would change as "there have been numerous research efforts" or something similar
%there has been lots of research
there have been numerous research efforts towards the development of various quality metrics and models which can help predict
%MM perhaps  predicting better but left as not 100% sure
the end-user QoE of the multimedia application as perceived by the end-user \cite{barman2019qoe}. Such quality metrics can vary from simple image quality metrics such as PSNR and SSIM~\cite{SSIM} to more complex video quality metrics such as VMAF~\cite{NetflixVMAF_Github} and ITU-T Rec. P.1204~\cite{1204}.

One of the reasons behind the advancement of the field of QoE, such as improved QoE models and metrics and QoE-based optimization of video streaming workflow, is due to the creation and availability of open-source datasets, from datasets from the early 2000s, such as VQEG-HD3~\cite{cdvl} and Live VQA~\cite{LIVEVideoQualityAssessmentDatabase}, to more recent datasets such as AVT-VQDB-UHD1~\cite{rao2019AVT-VQDB-UHD-1}, GamingVideoSET~\cite{GamingVideoSET}, Live YouTube-Gaming~\cite{LiveYouTubeGaming} and BC-KU MultiScreen Dataset~\cite{barmanMultiScreen}. 

\subsection{Motivation}
Over the years, many subjective and objective quality assessment datasets have been created and made available to the community 
%over the years
~\cite{GamingVideoSET,rao2019AVT-VQDB-UHD-1,EPFL-Polimi_Website,LIVEVideoQualityAssessmentDatabase,LIVEQoEDatabaseforHTTPbasedVideoStreaming,LiveYouTubeGaming,barmanMultiScreen,barman2022lcevc,LiveMobileStallVideoDatabaseII,LIVENetflixVideoQualityofExperienceDatabase,LIVE-NFLX-IISubjectiveVideoQoEDatabase,LFOVIAVideoQoEDatabase}. In order to streamline and standardize the process of conducting subjective tests and objective quality (model) evaluation, various standards such as ITU-T P.808~\cite{p808}, ITU-T P.809~\cite{p809}, ITU-T P.910~\cite{p910}, ITU-T P.913~\cite{p913}, ITU-R BT.500~\cite{bt500-14}, and ITU-T P.1401~\cite{P1401} have been proposed. Such standards provide detailed recommendations on various aspects, such as the selection of video sequences, subjective test procedure (test environment, participant selection, test methodology, etc.), and model performance evaluation. 

However, there is no ``standard”/template that outlines the documentation process to describe the various aspects of the dataset. In the absence of such templates, it is left to the creators of the dataset to report the various aspects of the dataset. The absence of a template can inadvertently result in the omission of important information about the dataset. This also shifts the onus of gathering and documenting the information from relevant publication(s) to the end-user, which is time-consuming, tedious, and often 
non-reproducible.

\subsection{Prior Work/Efforts}

The need for documenting datasets is not exclusive to QoE datasets. For example,  data provenance has been studied extensively in other fields, such as in the databases community~\cite{datahub,provenance}. Similarly, more recently, many works have focussed on the process of documenting the creation and use of machine learning datasets. Examples of such works include model cards~\cite{Mitchell2019ModelCards} and datasheets~\cite{Gebru2021Datasheet}, which allow the dataset creators to document various aspects of machine learning models and datasets. Such works have found good acceptance in the machine learning community due to their high utility in enhancing the communication and transparency between the dataset creator and users. 

This paper is inspired by the work of Gebru \textit{et al.}~\cite{Gebru2021Datasheet}, where the authors have presented a datasheet for AI/ML-based datasets. However, the proposed datasheet template (and other similar works) are unsuitable for QoE datasets as they are designed for typical AI/ML-based datasets, which are usually huge (millions of images/billions of text, 100 thousands of videos) and have attributes focussing on the model development process. QoE datasets, on the other hand, are typically much smaller and focus on the subjective and objective assessment of audiovisual content. 

\subsection{Contributions}

This paper presents a ``\textit{datasheet}” template to document various aspects of QoE (subjective and objective assessment) datasets. The proposed datasheet consists of various questions/aspects grouped into six different sections. Each individual field/question is supported with a detailed description. The datasheet can be filled by either the dataset creator or the end-user and then shared for easier understanding and reproducibility of their work. 

The datasheet template is publicly available on GitHub~\cite{datasheet_qoe_github} in various formats (\textit{google sheets}, \textit{.xlsx}, \textit{.odt}, \textit{.pdf}, and \textit{.html}) and can be used to document the various aspects of both new and existing datasets. Along with the proposed datasheet template, for easier understanding, various example datasheets of existing open-source datasets have also been made available in the GitHub repository~\cite{datasheet_qoe_github}. 

\subsection{Template Development Process}

We elaborate in this section on the datasheet template's creation process. The authors first created the draft datasheet template based on their extensive experience in the field of QoE assessment. More specifically, learnings while creating and documenting
% MM - I think about 10 considering those with Thushara, the one with Roopak/Peter, the one with Moustafa.
over ten open-source datasets, along with their experience in using similar third-party open-source datasets for various purposes, were used to design the draft template. The template was then used to create example datasheets for three open-source datasets, GamingVideoSET~\cite{GamingVideoSET}, AVT-VQDB-UHD-1~\cite{rao2019AVT-VQDB-UHD-1} and BC-KU Multi-Screen dataset~\cite{multiscreen_github}. Based on our experience filling in the example datasheet, we identified the missing and wrongly ordered fields, typos, and lack of clarity in titles and descriptions, which was incorporated into improving the draft datasheet template. After this, feedback was then collected from QoE researchers from Sony (Germany), TU Illmenau (Germany), and Kingston University (UK), which was then used to further improve the datasheet to obtain the current proposed version of the datasheet template. 

%######################## start of the figure ####################################
\begin{figure}[t!]
\begin{center}
    \includegraphics[trim=0cm 3cm 7cm 0,clip,width=1.05\linewidth]{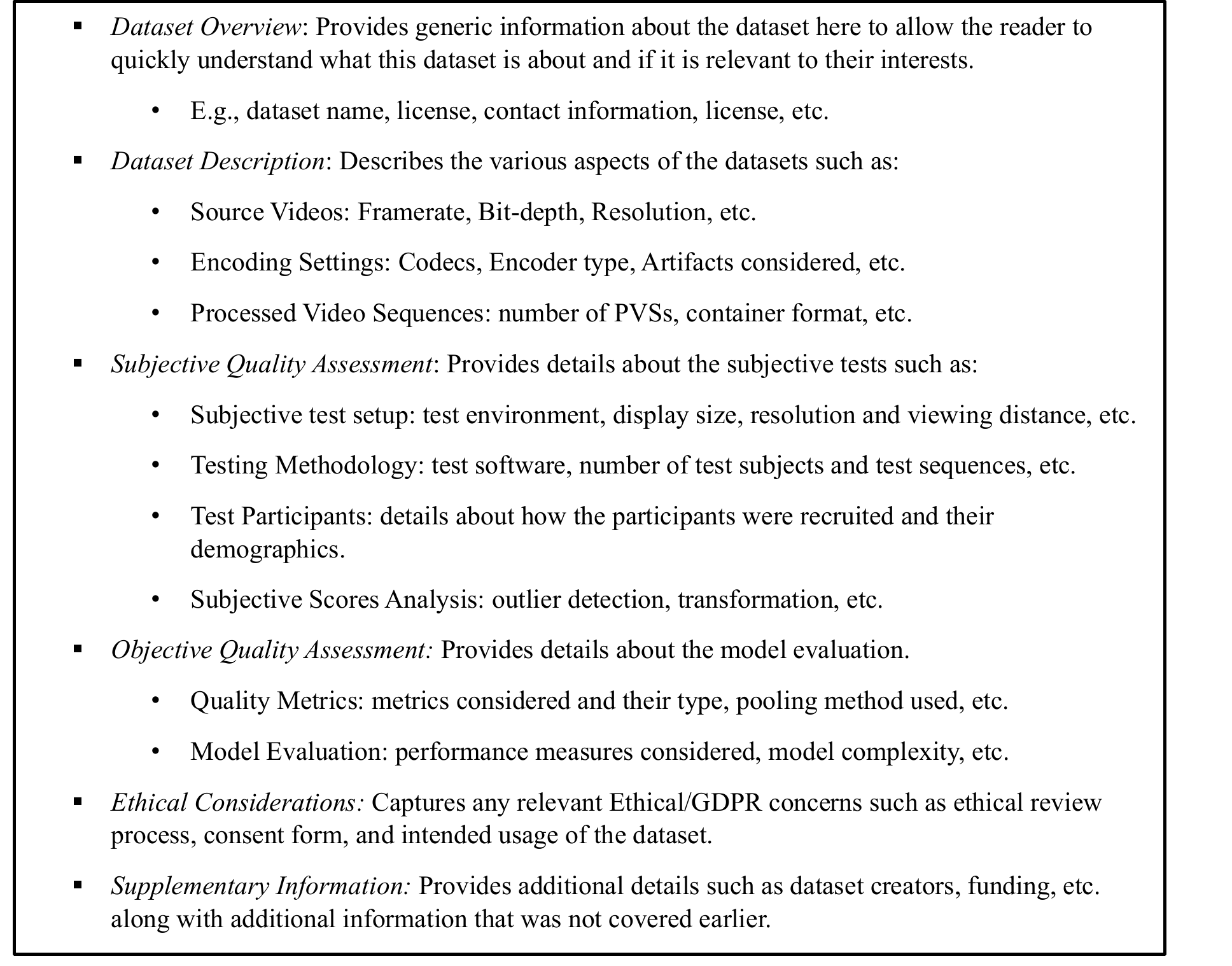}
% Answer: [trim={left bottom right top},clip]
\end{center}
\caption{Summary of various sections of the proposed datasheet template.}
\label{fig:summary}
\end{figure}
%######################## end of the figure ####################################

%########################################################################################
\section{Proposed Datasheet Template}

Figure~\ref{fig:summary} presents a summary of the proposed datasheet template. The datasheet considers various aspects of any traditional 2D audiovisual QoE datasets, from dataset overview to details about subjective and objective quality assessment. The proposed fields are optional and provide the flexibility to add any additional information as the dataset creator desires. For easier understanding, the proposed datasheet is divided into six different sections, as discussed next.

\subsection{Dataset Overview}

As the name suggests, this first section provides an overview of the dataset to the end user. It is intended to allow the readers to quickly understand what the dataset is about and if it is relevant to their interests. Details such as the name of the dataset, the date/year it was created, the dataset repository download link, license, required citation, and contact information are presented in this section. 

\subsection{Dataset Description}

The second section summarizes the various characteristics of the dataset's source and encoded video representations and is further divided into three sub-sections. In the first part, ``Source Videos", information about the various aspects of the source videos (number and type, bit-depth, dynamic range, resolution(s), etc.) that are either provided or used in the dataset are captured. This is quite important as the reader might be interested in a dataset with a particular type of content (e.g., 10-bit HDR gaming content). In the second part, ``Encoding Settings," information about the encoding parameters, such as encoder type, rate control, codecs, resolution, bit-depth, etc., is collected. The last part, ``Processed Video Sequences," captures details about the encoded video sequences, such as the number of sequences used and the container format used for media playback. 

\subsection{Subjective Quality Assessment}

A detailed description of the test settings, methodology, and procedures that must be followed, including data processing guidelines, such as outlier detection, etc., as defined in various ITU Recommendations~\cite{p910,p913,bt500-14} can help in asserting the reliability, repeatability, and validity of the reported subjective test results. Hence, this section captures the relevant information covering various aspects of subjective quality assessment: subjective test setup (test environment, display, viewing distance, rating scale, etc.), 
%MM - methodology is repeated, if you were referring to different aspects you could mention it here.
testing methodology (playback software, number of test subjects and sequences, etc.), information about test participants (demographics of test participants, etc.), and subjective data analysis (outlier analysis, score transformation, etc.). The information presented in this section can help the reader better understand the subjective test assessment results. 

\subsection{Objective Quality Assessment}

Objective Quality Assessment includes methods and models that use objective measurements such as signal fidelity to predict the visual quality as perceived by human observers. Most QoE datasets include the performance evaluation of various image and video quality metrics. This section collects data that tries to capture the different aspects of the model performance evaluation, such as the quality metrics considered, the implementation used and its version, how the model was trained and tested, and various measures that were used to quantify the performance of evaluated models and metrics.

\subsection{Ethical Considerations}

QoE datasets often include the use of multimedia data, which can include personal data such as images/videos of persons or, in some cases, might include violent/disturbing scenes. For example, in the case of datasets containing gaming videos, there might be violent scenes that some viewers might find disturbing. Also, often QoE datasets include a subjective quality assessment that includes human test subjects. Hence, it is often of significant importance to the dataset users to understand the relevant ethical approval/considerations that were taken into account during the design of the dataset. Various aspects, such as the intended usage of the dataset, ethics approval, sample consent form used during the subjective tests, and other relevant ethical or GDPR concerns the reader must be aware of before using this dataset, are collected in this section. 

\subsection{Supplementary Information}

This section includes additional questions that try to capture information that does not form an integral part of the rest of the sections. This includes information about the creators/authors, how the dataset creation was funded, and any confidential aspects of the dataset the reader needs to understand. Additionally, this section provides the opportunity to add any other relevant supplementary information by the dataset creator that was not captured by the various questions in the template. 

\noindent \textit{Note:} It should be noted that the proposed template includes much more features than what has been discussed above. Please refer to the actual datasheet template in GitHub\cite{datasheet_qoe_github} or Appendix at the end of this paper for a complete overview of the proposed datasheet. 

\subsection{Example Datasheets}
In order to help the reader better understand the proposed template, we provide example templates for the following datasets:
\begin{enumerate}
    \item GamingVideoSET (2018)~\cite{GamingVideoSET}: This dataset consists of source videos and subjective and objective assessment results for gaming video quality assessment. 
    \item AVT-VQDB-UHD1 (2020)~\cite{rao2019AVT-VQDB-UHD-1}: Dataset consisting of source videos, subjective and objective scores for videos encoded with three different codecs, which was in part used in the design of ITU-T Rec. P.1204~\cite{1204}.
    \item BC-KU Multi-Screen Dataset (2023)~\cite{barmanMultiScreen}: A very recent dataset consisting of subjective and objective assessment results considering a multiscreen setup of three different devices: mobile, tablet, and TV.
\end{enumerate}

%########################################################################################
\section{Discussion, Conclusion and Future Work}

We presented in this paper a first attempt at creating a datasheet template to enable better documentation of subjective and objective quality assessment datasets. It is not definitive or complete, and we anticipate continuously improving it over time, based on discussions and feedback from other experts. As of the writing of this paper, additional feedback is being sought from experienced QoE researchers from various multimedia streaming companies and organizations such as Video Quality Experts Group  (VQEG)\footnote{https://vqeg.org/vqeg-home/} and Qualinet\footnote{http://www.qualinet.eu/}. The discussions and input will then be incorporated into developing an improved version which will be updated in the GitHub repository. We believe that this datasheet template can help both dataset creators, dataset users, and, where applicable, the reviewers of papers/works published based on the datasets.

While the requirement to fill in this datasheet does add overhead to the dataset creators, we believe that the benefits far outweigh the costs, as has also been the case in the field of AI/ML. Also, while the proposed template is more suitable to traditional 2D video/audiovisual datasets, it can easily be adapted to other datasets, such as Audio-only, Immersive Video (VR/AR, 360, Light Field, Point Cloud and 3D meshes) 
%MM - Light field is a way to capture and visualise immersive video hence I would include it before. Same for point cloud and 3D meshes. 
and Computer Vision. In the future, collaboratively with other QoE experts, we plan to create more personalized datasheets for such QoE datasets. 

%########################################################################################

\section*{Acknowledgments}
Nabajeet Barman would like to thank Rakesh Rao, Saman Zadtootaghaj, and Steven Schmidt for their feedback on the draft template. 
%#############`###########################################################################

\begingroup
\setstretch{0.85}
\printbibliography
\endgroup

\appendix
For an easier understanding of the template, an example datasheet for an open-source dataset GamingVideoSET is provided here. The datasheet template and all example datasheets can be found in the GitHub repository~\cite{datasheet_qoe_github}.


\section*{Supplementary material}

**Datasheet for QoE Datasets.** This template is intended to be used by both dataset creators as well as anyone using an already published dataset.

*Note: If there are multiple datasets that are being used/proposed in a single work, please fill in a separate sheet for each individual dataset*

| **Template Details** (not to be modified) | Date Created: | 28-April-2023 | Date this template was first made publicly available |
| --- | --- | --- | --- |
| | Date Modified | NA | Date the template was modified, if applicable |
| | Version | v 1.0 | Current Version Number |

| **Section I: Dataset Overview** | | **Value** | **Additional Comments/Urls (as applicable)** | **Description of the field** |
| --- | --- | --- | --- | --- |
| Provide generic information about your dataset here. It will allow the reader to quickly understand what this dataset is about and if it is relevant to their interests. | Dataset Name | GamingVideoSET | | Name of the dataset |
| | Dataset Abbreviation | NA | | Short form of the dataset name, if different from Dataset name |
| | Version | v1.0 | | Version (default: v1.0) |
| | Creation/Publication Date/Year | 2018 | | Date the dataset was first published. If exact date not available, please add the year of creation |
| | Modification/Update Date (dd-mm-yyyy) | NA | | Date the dataset was last updated (write NA if not applicable/updated since creation) |
| | Repository | https://kingston.box.com/v/GamingVideoSET | | Link to the repository |
| | Citation (BibTex/PlainTex/Url) | N. Barman, S. Zadtootaghaj, S. Schmidt, M. G. Martini and S. Möller, "GamingVideoSET: A Dataset for Gaming Video Streaming Applications," 2018 16th Annual Workshop on Network and Systems Support for Games (NetGames), Amsterdam, Netherlands, 2018, pp. 1-6. | | Please add the required citation(s) or point to an url providing the citation |
| | License | NA | | MIT, Apache 2.0, BSD, etc. |
| | Open-Access? | Yes | Password protected | Is the dataset fully open or password protected? |
| | Contact Information | Nabajeet Barman (nabajeetbarman4@gmail.com) | | Contact information of the authors/creators to send questions or comments about the dataset |
| | Dataset Size | 30 GB | | (Approx) size of the full dataset (in GB) |
| | Additional Information? | NA | | Add any additional information, if available |

| **Section II: Dataset Description** | | **Value** | **Additional Comments/Urls (as applicable)** | **Description of the field** |
| --- | --- | --- | --- | --- |
| **Source Videos** Information about the various aspects about the source videos that are used in the study | Number of Source Sequences | 24 | | Number of source sequences considered (audio, videos, etc.) |
| | Content Genre | Gaming | | Nature of source sequences (gaming, natural, animation, computer generated, etc.) |
| | Source Sequence Available? | Yes | | Are source sequences made available as part of the dataset? |
| | Source Data Repository(ies) | NA | | Please write NA if the dataset includes source videos as a contribution. |
| | Bit-depth (s) | 8 | | 8-bit, 10-bit, 12-bit, 16-bit, etc. |
| | Dynamic Range | SDR | | SDR/HDR (please add details where possible, e.g., transfer char, gamma, etc.) |
| | Frame rate(s) | 30 | | Frame rate of the source videos, e.g., 24, 25, 30, 60, 120, 144, etc. |
| | Resolution(s) | 1920x1080 | | Resolution of source videos |
| | Pristine/User Generated | Pristine | Captured losslessly | Are the source sequences pristine or user generated/already compressed? |
| | Video Format | Raw Video | | Format of the source videos, e.g., RawVideo, ProRes, HEVC Encoded, etc. |
| | Video Container | YUV | | MP4, MKV, WEBM, Y4M |
| | Audio Format (if applicable) | NA | | Audio format, if available |
| | Video/Audio/Audiovisual | Video only | | Type of source sequences (e.g.,video only, audio only, audiovisual) |
| | SI/TI Information | Yes | | Is SI/TI [ITU-T Rec. 913] available in the dataset? |
| | Additional Information? | None | | Add any additional information, if available |
| **Encoding Settings** Information about the encoding settings used in the dataset (also referred to as Hypothetical Reference Circuit (HRC)) in some works). | Encoder Implementation | FFmpeg | | e.g., FFmpeg, VTM, HM, etc. |
| | Encoder Type | Software | | Software/Hardware/Both |
| | Type of Software implementation | Practical | | Reference/Practical/Both |
| | Rate Control | CBR | | VBV, CBR, ABR, etc. |
| | Encoding Speed/Mode | veryfast | | e.g., FFmpeg: ultrafast, medium, slow, etc. |
| | Video Codec(s) | H.264 | Profile Main 4.0 | H.264, HEVC, VP9, AV1, VVC, AVS2, AVS3 |
| | Resolution(s) | 1080p, 7200, 480p | | Encoding resolutions considered |
| | Framerate(s) | 30 | | Encoding framerates considered |
| | Bit-depth | 8 | | Encoded (video) sequences bitrates |
| | Artifacts Considered | compression and scaling | | Encoding, Scaling, Network (Stalling, Quality Changes, Packet Loss, etc.). Please add details |
| | Scenario considered | HAS Live Streaming | | Type of streaming scenarios considered |
| | Additional Information? | NA | | Add any additional information, if available |
| **Processed Video Sequences** | Number of PVSs | 576 | | How many PVSs are in total? |
| | Container Format | MP4 | | MP4, MKV, WEBM, Decoded Rawvideo (YUV), etc. |
| | Additional Information | None | | Add any additional information, if available |

| **Section III: Subjective Quality Assesment** | | **Value** | **Additional Comments/Urls (as applicable)** | **Description of the field** |
| --- | --- | --- | --- | --- |
| **Subjective Test Setup** | Subjective Test Environment | Lab-based | | Controlled (Lab), Public, Home-based, Crowdsourcing, Hybrid |
| | ITU Rec | BT.500-14 | | Please mention if the test followed any specific ITU Rec such as BT.500-14, P.910, P.913 |
| | Subjective test methods and rating scales | ACR(1-5) | | ACR(1-5), ACR(0-10), DSIS, CCR? If not standardized method, please describe in details. |
| | Rating Scale | Overall | | Overall / Continuous (using a slider) / Both |
| | Display Type | Desktop Monitor | | TV, Mobile, Tablet, Desktop Monitor |
| | Display Size | 24" | ViewSonic display monitor | Size of the display(s) used |
| | Display Resolution | 1920x1080 | | Link to the device model, if available |
| | Viewing Distance | 3H | | Viewing distance from the screen |
| | Viewing Angle | Yes | In paper | |
| | Subjective test schematic or photo | NA | | Is there any schematic or photo showing the lab test set up? If yes, please mention where it is available? |
| | Light Intensity in Test Room | NA | | Was device, room illuminance, gamma level, etc. measured? If yes, please add details |
| | Additional Information | NA | | Add any additional information about the test environment, if available |

| Category | Field | Value | Additional Comments/Urls (as applicable) | Description of the field |
|---|---|---|---|---|
| **Test Methodology** | Video Playback Software | VLC | | Video player used to play the videos (e.g., VLC, MPV, Matlab, FFplay, etc.) |
| | Test Software | Proprietary Software | | Which software did you use to play the videos and record the subjective opinion scores? E.g., Matlab, MSU VQMT, TUIL AVRateNg, Subjectify.us, Proprietary, Paper-based. Please add reference, as applicable |
| | Training | Yes | using other gaming sequences | Was there any training session before the actual test? |
| | Playlist randomized | Yes | | Was the playlist randomized for each session/participant? Please explain. |
| | Playback | Self-paced by participant | | How was the stimuli playback performed? E.g., fixed duration video followed by grey screen for a fixed duration, self-paced by the test participant. |
| | Upscaling (if used) | Bicubic | | Upscaling filter used (Bicubic, Lanzcos-3, Lanzcos-5, Super Resolution) |
| | Test type | Video only | | Video only, Audio-visual, audio only |
| | Number of Test Subjects | 24 | | Number of participants that took part in the subjective test |
| | Test subjects per session | 1 | | How many test subjects took part in a single session? |
| | Number of Test Sequences | 90 | | Number of test sequences considered in subjective test |
| | Additional Information | NA | | Add any additional information about testing methodology, if available |
| **Test Participants** | Test subject recruitment | University students and staff | | How were the test subjects recruited for the study? University students and staff, Crowdsourcing platforms, friends and family, etc. |
| | Pre-screening? | Both performed in lab | | Was there any eye test or any othe pre-screening test performed on the subjective test participants? if yes, please mention which ones (visual acuity test such as Sellen or color-blindness test using Ishira charts, 3D vision check using stereo butterfly test, etc.) |
| | Experts or Non Experts | Non-experts | | Please mention if the test subjects are experts (people working in video streaming) or non-experts or mix of both? |
| | Gender Distribution | 32% Male, 68% Female | | What is the gender distribution of the test subjects? |
| | Age Distribution | (Median) 29 | | What is the age distribution of the test participants? |
| | Additional Demographic Information | Yes | | Is additional demographic information data such as viewing device preferences, number of hours videos streaming, gaming experience, etc. (as applicable) available? If yes, please provide details |
| | Additional Information | NA | | Add any additional information about test participants, if available |
| **Subjective Scores Analysis** | Post-Screening | Yes | IQR based, please see paper for details | Did you perform any outlier analysis? If yes, which methodology did you use? Eg., Correlation based, Z-score |
| | Subjective Scores Removal? | Yes | | Were any subjective scores removed after outlier analysis? |
| | Subjective Score(s) | MOS | | How are the subjective scores reported? MOS/DMOS only? |
| | Subjective Scores Transformation | Yes | Please see Ref-Pezulli | Did you perform any mapping or transformation on the raw subjective scores? If yes, which one (e.g., generalized linear models (GLMs) for estimation of the population average QoE [Ref-Pezulli]. Please add details? |
| | Individual subjective scores | Yes, as part of additional dataset | see Note 1 below | Are individual subjective scores available? |
| | Statistical Analysis? | No | | Was there any statistical analysis performed such as calculation of CI, etc. |
| | Additional Information | NA | | Add any additional information about subjective scores post-processing and analysis, if available |
| **Any other comments** | | None | | Add any additional information about the subjective test which is not covered above and might be relevant to the reader |

| **Section IV: Objective Quality Assessment** | | Value | Additional Comments/Urls (as applicable) | Description of the field |
|---|---|---|---|---|
| **Quality Metrics** | Quality Metrics Considered | PSNR, SSIM, VMAF | | Which quality metrics are you evaluating in this work? e.g., PSNR, SSIM, VMAF, LPIPS, |
| | Metric Type | Traditional, Machine Learning | | What are the type of metrics that you are using? ML/DL/Traditional, Mixed |
| | Pooling | VQMT tool | See Note 2 | What type of pooling method (e.g., mean, minkowski summation, harmonic mean, etc.) is used to obtain the final video quality score (if applicable). |
| | Implementation(s) and Version | Only metric scores were made available in the dataset. | | Add the list of urls to the open-source implementation links for the metrics used, if available (ffmpeg,github, commercial, proprietary, etc.). Please also add the version of the metric and/or implementation that you have used in this work |
| | Additional Information? | NA | | Add any additional information about the quality metrics that might be relevant to the user |
| **Model Evaluation** | Performance Measures Considered | PLCC and SROCC | | Which measures are considered to evaluate the performance of the quality metrics (e.g., PLCC, SROCC, RMSE, Kendall, R2, etc.) |
| | Mapping performed | None | | Is there any transformation/fitting (linear/3rd order polynoimial/logistic regression) performed before the computation of the performance scores? |
| | Statistical Significance Test(s) | Yes | See Note 2 | Was any statistical significance test performed to compare the performance of the metrics compared/evaluated? |
| | Model Training | NA | | Was the model(s) trained/retrained? If yes, how? |
| | Model Complexity | NA | | If available, please add how the model complexity was evaluated |
| | Additional Information? | NA | | Add any additional information about model evaluation that might be relevant to the user |
| **Any other comments** | | None | | Please add any additional information about the objective evaluation which is not covered above and might be relevant to the reader |

| **Section V: Ethical Considerations** | | Value | Additional Comments/Urls (as applicable) | Description of the field |
|---|---|---|---|---|
| Please use this section to describe any Ethical/GDPR related information | Ethical review | Yes | TU Berlin | Were any ethical review processes conducted before the dataset creation? |
| | Intended Use | Non-Commercial Research Only | | Please describe the intended use of this dataset? |
| | Consent Form (used for subjective tests) | NA | | Link to the sample consent form that was provided to the test subjects before their participation, if available |
| | Ethical/GDPR concerns? | Some sequences might contain violent scenes | Please refer to the readme file in the dataset | Are there any ethical concerns? that might be relevant to the reader/user of this dataset |
| | Additional Comments? | NA | | Is there any additional ethical considerations/information that would be relevant to the users of this dataset? |

| **Section VI: Supplementary Information** | | Value | Additional Comments/Urls (as applicable) | Description of the field |
|---|---|---|---|---|
| | Are preview/web version of source videos available? | No | | Is there any preview/sample sequences available for someone to view/listen to understand the dataset? This is useful when the dataset is of huge size and the reader would like to make sure that it suits their requirements before downloading the same. |

| Answer the additional questions mentioned here, as applicable. If there is any other information that you would like to add, please add it at the end of these questions | Who created the dataset? | Academic Collaboation | Kingston University, London and Technical University, Berlin, Germany | Company, University, Collaboration |
|---|---|---|---|---|
| | Who funded the creation of the dataset? | EU and DFG | EU Horizon 2020 grant agreement No 643072 and DFG Project MO 1038/21-1. | Project, company funding, self-funded, no funding |
| | Is there anything in the dataset that is confidential or might restrict its usage? | Fair Usage Policy Applies | Please refer to the readme file in the dataset | For example, is the dataset meant only for research use and non-commercial usage |
| | Who performed the subjective tests? | Authors | | Was the subjective test performed by the authors/creators of the dataset? Or was it outsourced to an external 3rd-party? Please provide details |
| | Existing Works where the dataset is being used | Please see references [3-10] | Others can be found on google scholar | Please add references to existing work(s) where either you or someone else have used this dataset (this is applicable to existing datasets that might have been available for sometime) |

| Please use this space to add any additional information that you think might be helpful to the reader which was not covered in this datasheet questionnaire | Note 1 | More detailed analysis into VQA performance was perfromed in a subsequent publication. Please see Ref [1]. | Additional Comments/Information |
|---|---|---|---|
| | Note 2 | VQMT Tool available in [2] was used for metric computation | Additional Comments/Information |
| | Note 3 | | Additional Comments/Information |
| | Note 4 | | Additional Comments/Information |
| | Note 5 | | Additional Comments/Information |
| | Note 6 | | Additional Comments/Information |

Where not explicitly mentioned, this proposed template uses terminologies as defined in ITU-T P.10/G.100 and in other related ITU-T Rec (P.910, P.913, P.1401) and ITU-R BT.500

**End of the Datasheet**



\end{document}